\documentclass[reprint, showpacs]{revtex4-1}
\usepackage{graphicx}
\usepackage{amsfonts}
\usepackage{natbib}
\usepackage{float}

\newcommand{\be}{\begin{equation}}
\newcommand{\ee}{\end{equation}}
\newcommand{\bea}{\begin{eqnarray}}
\newcommand{\eea}{\end{eqnarray}}

\newcommand{\al}{\alpha}

\newcommand{\mA}{\left< A \right>}
\newcommand{\sA}{\left< A^2 \right>}

\begin{document}
\title{Effects of stickiness in the classical and quantum ergodic lemon billiard}

\author{\v Crt Lozej}
\author{Dragan Lukman}
\author{Marko Robnik}

\affiliation{CAMTP - Center for Applied Mathematics and Theoretical
  Physics, University of Maribor, Mladinska 3, SI-2000 Maribor, Slovenia, European Union}


\date{\today}

\begin{abstract}
  We study the classical and quantum ergodic lemon billiard introduced by
  Heller and Tomsovic in Phys. Today {\bf 46} 38 (1993),
  for the case B=1/2, which is a classically
  ergodic system (without a rigorous proof) exhibiting strong stickiness
  regions around a zero-measure bouncing ball modes. The structure of the
  classical stickiness regions is uncovered in the S-plots introduced by
  Lozej in Phys. Rev. E {\bf 101} 052204 (2020). A unique classical transport
  or diffusion time cannot be defined. As a consequence the quantum states
  are characterized by the following {\em nonuniversal} properties:
  (i) All eigenstates are
  chaotic but localized as exhibited in the Poincar\'e-Husimi (PH) functions.
  (ii) The entropy localization measure A (also the normalized inverse
  participation ratio) has a nonuniversal distribution, typically bimodal,
  thus deviating from the beta distribution, the latter one being
  characteristic of uniformly chaotic systems with no stickiness regions.
  (iii) The energy level spacing
  distribution is Berry-Robnik-Brody (BRB), capturing two effects:
  the {\em quantally divided phase space} (because most of the PH functions are
  either the inner-ones or the outer-ones, dictated by the classical stickiness,
  with an effective parameter $\mu_1$ measuring the size of the inner
  region bordered by the sticky invariant object, namely a cantorus),
  and the {\em localization of PH functions}
  characterized by the level repulsion (Brody) parameter $\beta$.
  (iv) In the energy range considered (between 20.000 states to 400.000 states
  above the ground state) the picture (the structure of the eigenstates
  and the statistics of the energy spectra)  is not changing qualitatively,
  as $\beta$ fluctuates around $0.8$, while $\mu_1$ decreases almost
  monotonically, with increasing energy. 
\end{abstract}

\pacs{01.55.+b, 02.50.Cw, 02.60.Cb, 05.45.Pq, 05.45.Mt}

\maketitle

\section{Introduction}
\label{sec1}

Quantum chaos, or more generally wave chaos, is an established field of
research in physics \cite{Stoe, Haake, Rob2016}. The existence of dynamical
chaos in quantum mechanics is still a subject of current debates. 
The sensitive dependence of time evolution (solution of time dependent
Schr\"odinger equation) as an analogy of classical chaos certainly
does not exist, because the overlap of two initial states remains
rigorously constant due to the unitary time evolution.
Also, of course, the analogy of classical orbits does not exist
in quantum mechanics due to the Heisenberg uncertainty principle.
Usually the quantum systems have a classical correspondent.
If not, a classical correspondent can be constructed and studied
by means of introducing the coherent states, as is done - for example -
in the Dicke model (see Ref. \cite{WR2020} and references therein).  In
the following we refer to quantum systems that have classical Hamiltonian
correspondents.

The stationary quantum chaos is well established analogy of
the classical chaos in Hamiltonian systems \cite{Stoe, Haake, Rob2016}.
Namely, we find phenomena in the solutions of the time independent Schr\"odinger
equation which correspond exactly to the classical structures. Such
signatures of classical chaos are found in the statistical properties of
the energy spectra, in the structure of corresponding eigenfunctions and
of their Wigner functions \cite{Wig1932} or Husimi functions \cite{Hus1940}.
For example, the classically integrable systems exhibit Poisson statistics
of the unfolded (reduced to unit mean level spacing) energy spectra, their
wavefunctions have a well ordered structure of nodal lines or surfaces,
and their Wigner or Husimi functions are localized near the invariant
tori in the classical phase space. On the other hand, in the opposite
case of classically fully
chaotic (ergodic) systems the energy spectra obey the statistics of random
matrices, especially - but not only - of the Gaussian random matrices,
the nodal patterns of eigenfunctions are entirely disordered and
their probability amplitude exhibits a
Gaussian random function \cite{Berry1977}.
Their Wigner or Husimi functions are ergodic, in the sense
that they are on the average uniformly spread over the energy surface
in the classical phase space. For a review see Ref. \cite{Rob1998,Rob2016}.
If the classical limit does not exist the above criteria
can still be used as a definition of quantum chaos. 

The above statements are valid under an important {\em semiclassical condition},
namely that the dominating classical diffusion time or transport time
$t_T$ is sufficiently shorter than the Heisenberg time $t_H$, which by definition
is $t_H= 2\pi\hbar/ \Delta E$, where $\Delta E$ is the mean level
spacing, or inverse energy level density
$\rho(E)=1/\Delta E$.\footnote{It has been pointed out to us that historically
  this concept has been used explicitly for the first time by Victor Weisskopf,
  although in the literature we are unable to trace this back.}
In such case, if the semiclassical condition is satisfied,
all the above statements for fully chaotic systems
have been proven to be rigorously true using the semiclassical methods,
in particular Gutzwiller's semiclassical theory of expressing the
quantum Green function, and its trace $\rho(E)$, in terms of classical
periodic orbits
\cite{Gutzwiller1967,Gutzwiller1969,Gutzwiller1970,Gutzwiller1971,Gutzwiller1980}.
For fully chaotic systems, satisfying the
semiclassical condition, this proof was initiated by Berry \cite{Berry1985}
in 1985, further developed by  Sieber and Richter \cite{Sieber} in 2001,
and completed by the group of Haake
\cite{Mueller1,Mueller2,Mueller3,Mueller4} in the years 2004-2010 \cite{Haake}.
Therefore the well known Bohigas-Giannoni-Schmit conjecture \cite{BGS1984},
initiated by Casati, Valz-Gris and Guarneri \cite{Cas1980},
can be considered as proven.

Let us recall that the Heisenberg time goes to infinity when
$\hbar$ goes to zero, because $\Delta E \propto \hbar^{f}$ and
$f$ is the number of degrees of freedom, $f\ge 2$, as we
do not consider the systems having one degree of freedom. Thus,
in the semiclassical limit $\hbar \rightarrow 0$, the Heisenberg
time $t_H \propto \hbar^{1-f}$ eventually
becomes larger than any classical transport time of the system
$t_T$, as the latter one does not depend on $\hbar$. Their
ratio

\be \label{alpha}
\alpha = \frac{t_H}{t_T} =\frac{2\pi\hbar}{\Delta E \; t_T},
\ee
is the important parameter characterizing the deepness of the semiclassical
regime. Thus the semiclassical condition is $\alpha \gg 1$.
In such case the Principle of Uniform Semiclassical Condensation
(PUSC) \cite{Rob1998} of Wigner functions applies, saying that
the Wigner functions become uniformly spread over the classical
invariant component in the phase space, based on works by
Percival \cite{Percival1973}, Berry \cite{Berry1977}, Shnirelman \cite{Shnirelman1974},
Voros \cite{Voros1979}, and further developed by Veble, Robnik and Liu
\cite{VRL1999}. This can be an invariant torus, a chaotic component,
or the entire energy surface, depending on the dynamical properties and
the structure of phase space (integrable, mixed-type or ergodic).
Mixed-type systems
have been studied for the first time in the context of quantum chaos
by Berry and Robnik in 1984 \cite{BerRob1984}. Meanwhile the literature on
this problem has become quite extensive - for a recent review see Ref.
\cite{Rob2016,Rob2020}.

If the semiclassical condition $\alpha \ge 1$ is not satisfied,
we observe localization properties of the chaotic eigenstates uncovered in
the Wigner functions or Husimi functions in the phase space: The Wigner
or Husimi functions are concentrated on a proper subset of the available
classically chaotic region. In fact, the transition from strong
localization at $\alpha\ll 1$ to strong delocalization at $\alpha\gg 1$
is a rather smooth one, as observed recently in several model systems.
The chaotic regions in the classical Hamilton systems, either ergodic or
of the mixed-type, can have strongly nonuniform "chaoticity": there are
subregions that are more frequently visited by a chaotic orbit than the
others, and this difference can vary over orders of magnitude.
It can take a very long time
to exit such a sticky region, and symmetrically, a long time to enter, if
coming from outside. Such stickiness regions are bordered by cantori,
which are invariant remnants of destroyed inviariant tori, with fractal
dimension, and the size of their holes controls their permeability,
and therefore the classical transport time.
The quantification of the strength of the stickiness is characterized
in Sec. \ref{sec2} by the method of Lozej \cite{Lozej2020} in terms of
the so-called S-plots. For the literature on stickiness, introduced by
Contopoulos in 1971 \cite{Contopoulos1971}, see the review by Meiss
\cite{Meiss2015} and the references therein.
To study these effects in an ergodic billiard system with strong stickiness
is the main purpose of the present work, which follows a series of
our recent papers
\cite{BatRob2010,BatRob2013A,BatRob2013B,BLR2018,BLR2019B,BLR2020}.
As we shall see, stickiness implies
nonuniversal behavior of the statistics of the energy spectra and of the
localization measure.

In this paper we study a classically fully chaotic (ergodic) system,
namely a lemon billiard (B=1/2) introduced by Heller and Tomsovic in 1993
\cite{HelTom1993}, which possesses regions of strong stickiness, around the
zero-measure bouncing ball invariant component, making its chaoticity
strongly nonuniform. In this sense the system is nongeneric, not
of a mixed-type, but still exhibiting features which in the quantum
mechanics of the system imply nonuniversal, but very interesting
behavior. We are facing and studying the consequences of the stickiness
regions in the structure of eigenstates and of the corresponding
energy spectra. 

The paper is organized as follows. In Sec. \ref{sec2} we define the
family of lemon billiards, and describe its classical dynamical properties,
showing that due to the strong stickiness regions the system is
a nongeneric ergodic system. In Sec. \ref{sec3} we study the
corresponding quantum billiard  and define the Poincar\'e-Husimi
functions, and then explore their structure in correspondence with
the structure of the classical phase space. In Sec. \ref{sec4}
we define the entropy localization measure $A$ and the normalized
inverse participation ratio $R=nIPR$, showing that they are
approximately linearly related, and then explore statistical
poperties of $A$. In Sec. \ref{sec5} we analyze the statistical
properties of the energy spectra, showing that the level
spacing distribution is well described by the Berry-Robnik-Brody
distribution (BRB), because due to the classical stickiness effects
there is a region which quantum mechanics "sees" as an effective
regular island  whose relative size decreases almost monotonically
with increasing energy. The chaotic part of the spectrum is
subject to the Brody distribution with the level repulsion
parameter $\beta$ fluctuating
around 0.8 with changing energy.
In Sec. \ref{sec6} we present further comments
regarding the interpretation of the results. Sec. \ref{sec7}
presents the discussion, conclusions and outlook.

\section{The definition of the lemon billiard (B=1/2)
  and its classical dynamical properties}
\label{sec2}

\begin{figure}[H]
 \begin{centering}
    \includegraphics[width=9cm]{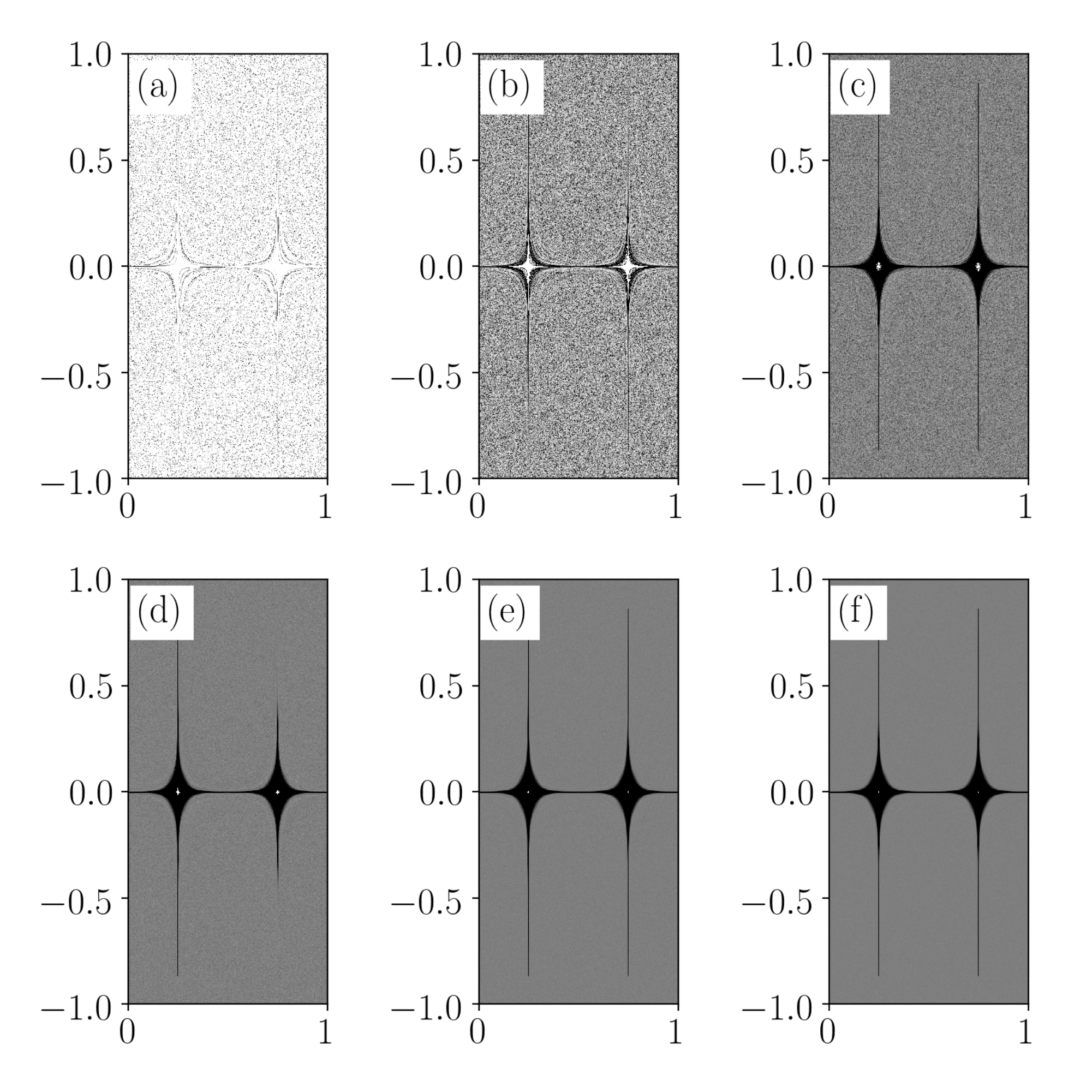}
   \par\end{centering}
 \caption{ The phase portrait as generated by a single chaotic
   orbit emanating from the same initial
   condition  $(s={\cal L}/2+0.0001,p=0.999)$
   after $10^6, 10^7, 10^8, 10^9, 5.10^9$, and $10^{10}$ collisions,
   from (a) to (f), respectively. The label on the abscissa is $s/{\cal L}$,
   while on the ordinate we have $p\in [-1,1]$.}
\label{fig1}
\end{figure}

The family of lemon billiards was introduced by Heller and Tomsovic
in 1993 \cite{HelTom1993}, and has been studied in a number of works
\cite{LMR1999,MHA2001,LMR2001,CMZZ2013,BZZ2016}, most recently
by Lozej \cite{Lozej2020} and Bunimovich et al \cite{BCPV2019}.
The lemon billiard boundary is defined
by the intersection of two circles of equal unit radius with the
distance between their center $2B$ being less than their diameters
and $B\in (0,1)$, and is given by the following implicit equations
in Cartesian coordinates

\bea   \label{lemonB}
(x+B)^2 + y^2 =1, \;\;\; x > 0, \\  \nonumber
(x-B)^2 + y^2 =1, \;\;\; x < 0.
\eea
As usual we use the canonical variables to specify the location $s$
and the momentum component $p$ on the boundary at the collision point.
Namely the arclength $s$ counting in the mathematical positive sense
(counterclokwise) from the point $(x,y) = (0, -\sqrt{1-B^2})$ as the
origin, while $p$ is equal to the sine of the reflection angle $\theta$,
thus  $p = \sin \theta \in [-1,1]$, as $\theta \in [-\pi/2,\pi/2]$.
The bounce map $(s,p) \Rightarrow (s',p')$ is
area preserving as in all billiard systems \cite{Berry1981}.
Due to the two kinks the Lazutkin
invariant tori (related to the boundary glancing orbits) 
do not exist. The period-2 orbit connecting the centers of the two
circular arcs at the positions $(1-B,0)$ and $(-1+B,0)$ is always
stable (and therefore surrounded by a regular island) except for
the case $B=1/2$, the subject of our present work, where it is
a marginaly unstable orbit (MUPO). One can see from the geometry that in
our case $B=1/2$ any orbit starting from middle of the circle
will hit the other circle at the right angle and thus will retrace
itself, because the centers exactly overlap with the arc of
the other circle. This is true for all angles $|\theta| \le \pi/3$ such
that $-p_0 \le p \le p_0$, where $p_0=\sqrt{3}/2 \approx 0.8660254$,
and the particle hits the corner of the lemon billiard. Therefore
we have a line of MUPO as an one-dimensional invariant object in the
phase space, namely located at $s={\cal L}/4$ and $s=3{\cal L}/4$
on the intervals $p\in (-p_0,p_0)$, where ${\cal L}$ is the circumference
of the entire billiard, for a general $B$ equal to

\be \label{perimeter}
{\cal L} = 4 \arctan \sqrt{B^{-2} -1}.
\ee
and for $B=1/2$ it is ${\cal L}= 4 \arctan \sqrt{3}= 4\pi/3 \approx  4.188790$.
The area ${\cal A} $ of the billiard for a general $B$ is equal to

\be \label{area}
{\cal A} = 2 \arctan \sqrt{B^{-2} -1} - 2B\sqrt{1-B^2},
\ee
thus for $B=1/2$ it is ${\cal A} = \frac{2\pi}{3} - \frac{\sqrt{3}}{2}
\approx 1.2283697$. Of course, correspondingly, we have also invariant
line of MUPO (we shall call them bouncing ball regions of period-4 orbits)
in the phase space located at $p=0$ and all $s$. It will be demonstrated
that these bouncing ball regions are surrounded by a very strong stickiness
region: even  after $10^{11}$ bounces of a single chaotic orbit,
starting at $(s={\cal L}/2+0.0001,p=0.999)$, there is still possibly a tiny
unoccupied island around $s={\cal L}/4$ and $s=3{\cal L}/4$ and $p=0$.
But we believe that the system is ergodic, lacking a
rigorous proof. The structure of the phase space
after $10^6, 10^7, 10^8, 10^9, 5.10^9$, and $10^{10}$ 
collisions emanating from the above initial
condition is shown in Fig. \ref{fig1}.

\begin{figure}[H]
 \begin{centering}
    \includegraphics[width=9cm]{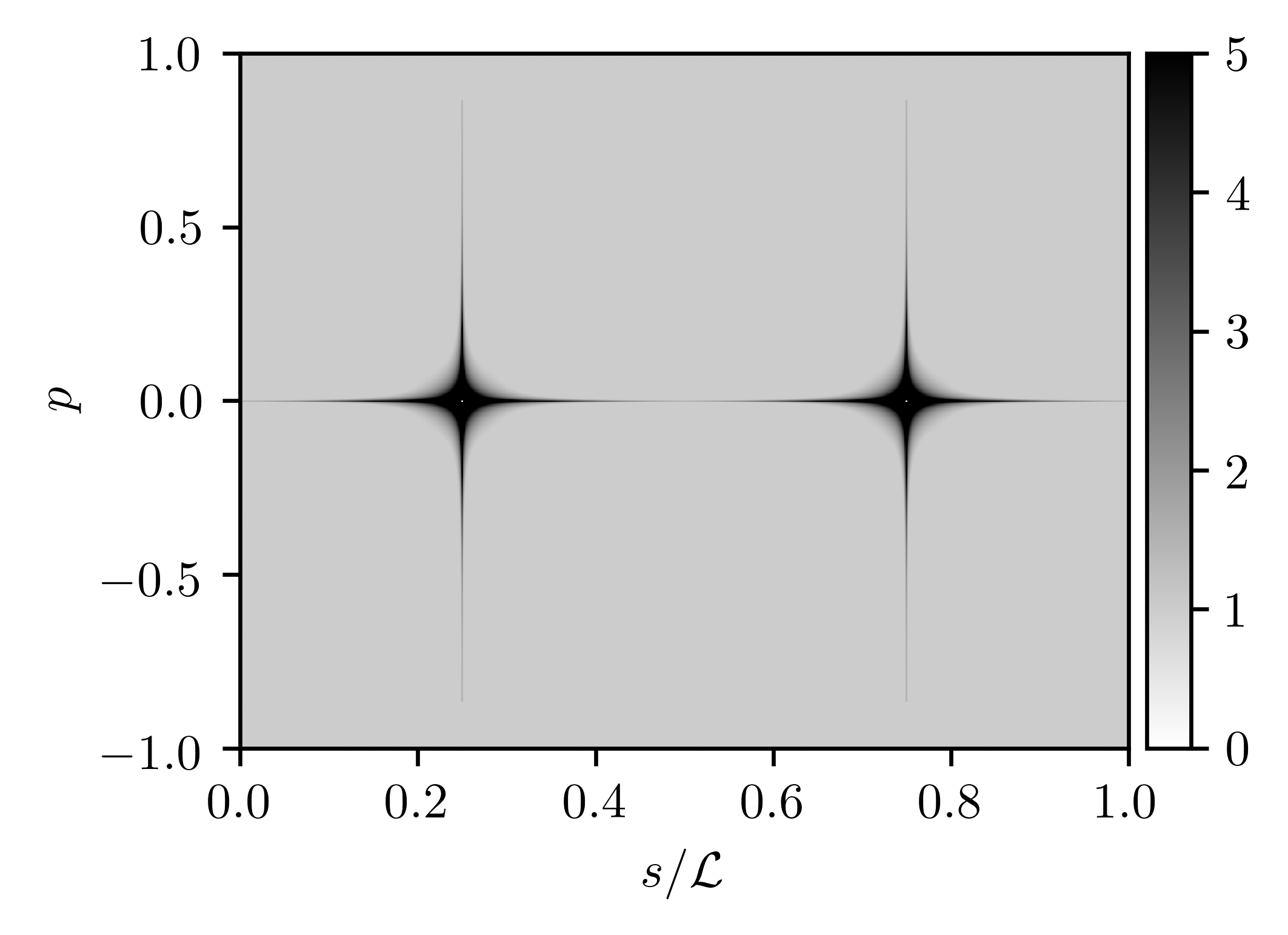}
   \par\end{centering}
 \caption{The S-plot on a grid of cells $1000\times 1000$,
   showing the extremely strong stickiness region around
 $s={\cal L}/4$ and $s=3{\cal L}/4$ and $p=0$.}
  \label{fig2}
\end{figure}

In Fig. \ref{fig2}
we show the S-plot using the method introduced in
\cite{Lozej2020}, to quantify the stickiness in the
phase space. In this approach the phase space is divided into
a network of equal cells defined by the uniform grid of
size $L$, in our case $L=1000$, implying $L^2=10^6$ cells,
and in each cell the distribution
of the discrete return times $\tau$ (number of iterations/bounces)
is observed, by calculating the
mean value of the return time $\langle \tau \rangle$
and of the standard deviation $\sigma$.  Their ratio
is the quantity $S=\sigma/\langle \tau \rangle$. If the
distribution of $\tau$ is Poissonian (exponential) characteristic
of uniform chaoticity described by the random model
\cite{Random}, we have $S=1$, while in the case
of stickiness we find distribution typically described by
the superposition of several exponential distributions
(so-called hyperexponential distribution) and $S > 1$.
Fig. \ref{fig2} clearly shows extremely strong stickiness in the
diamond shaped areas around $s={\cal L}/4$
and $s=3{\cal L}/4$ and $p=0$, which
has implications and manifestations in the quantum domain
to be studied in the next sections.

Lemon billiards of other values of $B$ are not considered
in this paper, but have been treated classically
in \cite{Lozej2020} and are subject of our forthcoming papers.

\section{The quantum billiard: The Helmholtz equation and
  the Poincar\'e-Husimi functions}
\label{sec3}

\subsection{The Helmholtz equation}
\label{sec3.1}

Quantum mechanically
we have to solve the stationary Schr\"odinger equation, which in a
billiard ${\cal B}$ is just the Helmholtz equation 

\be \label{Helmholtz}
\Delta \psi + k^2 \psi =0
\ee
with the Dirichlet boundary conditions  $\psi|_{\partial {\cal B}}=0$.
The energy is $E=k^2$. The important quantity is
the boundary function 

\be  \label{BF}
u(s) = {\bf n}\cdot \nabla_{{\bf r}} \psi \left({\bf r}(s)\right),
\ee
which is the normal derivative of the wavefunction $\psi$ at the 
point $s$ (${\bf n}$ is the unit outward normal vector). 
It satisfies the integral equation

\be \label{IEBF}
u(s) = -2 \oint dt\; u(t)\; {\bf n}\cdot\nabla_{{\bf r}} G({\bf r},{\bf r}(t)),
\ee
where $G({\bf r},{\bf r'}) = -\frac{i}{4} H_0^{(1)}(k|{\bf r}-{\bf r'}|)$ is
the Green function in terms of the Hankel function $H_0^{(1)}(x)$. It is important
to realize that the boundary function $u(s)$ contains complete information
about the wavefunction at any point ${\bf r}$ inside the billiard by the equation

\be \label{utopsi}
\psi_m({\bf r})  = - \oint dt\; u_m(t)\; G\left({\bf r},{\bf r}(t)\right).
\ee
Here $m$ is just the index (sequential quantum number) of the $m$-th eigenstate.

The number of energy levels ${\cal N} (E)$ below $E=k^2$ is determined quite
accurately, especially at large energies, asymptotically exact,
by the celebrated Weyl formula (with perimeter corrections)
using the Dirichlet boundary conditions, namely

\be \label{WeylN}
{\cal N} (E) = \frac{{\cal A}\;E}{4\pi} - \frac{{\cal L}\;\sqrt{E}}{4 \pi} +c.c.,
\ee
where $c.c.$ are small constants determined by the corners and the
curvature of the billiard boundary.  Thus the density of levels
$\rho (E) = d{\cal N}/dE$ is equal to

\be \label{Weylrho}
\rho(E) = \frac{{\cal A}}{4\pi} - \frac{{\cal L}}{8 \pi \sqrt{E}}.
\ee
Our numerical solving the Helmholtz equation is based on the plane wave
decomposition method and the Vergini-Saraceno scaling method
\cite{VerSar1995,LozejThesis}.
The numerical accuracy has been checked by the Weyl formula, to make sure
that we are neither losing levels nor getting too many due to the
double counting (distinguishing almost degenerate pairs from the
numerical pairs) in the overlapping energy intervals, and also by
the convergence test. The number of missing levels or too many levels was
never larger than 1 per 1000 levels (usually less than 10 per 10000 levels).

Our billiard has two reflection symmetries, thus four symmetry classes:
even-even, even-odd, odd-even and odd-odd.  For the purpose of analyzing
the spectral statistics we have thus considered only the quarter billiard,
while for the wavefunctions (and the corresponding PH functions)
we have used the half billiard of odd symmetry. 

\subsection{The Poincar\'e-Husimi functions}
\label{subsec3.2}

\begin{figure*}
 \begin{centering}
    \includegraphics[width=1\textwidth]{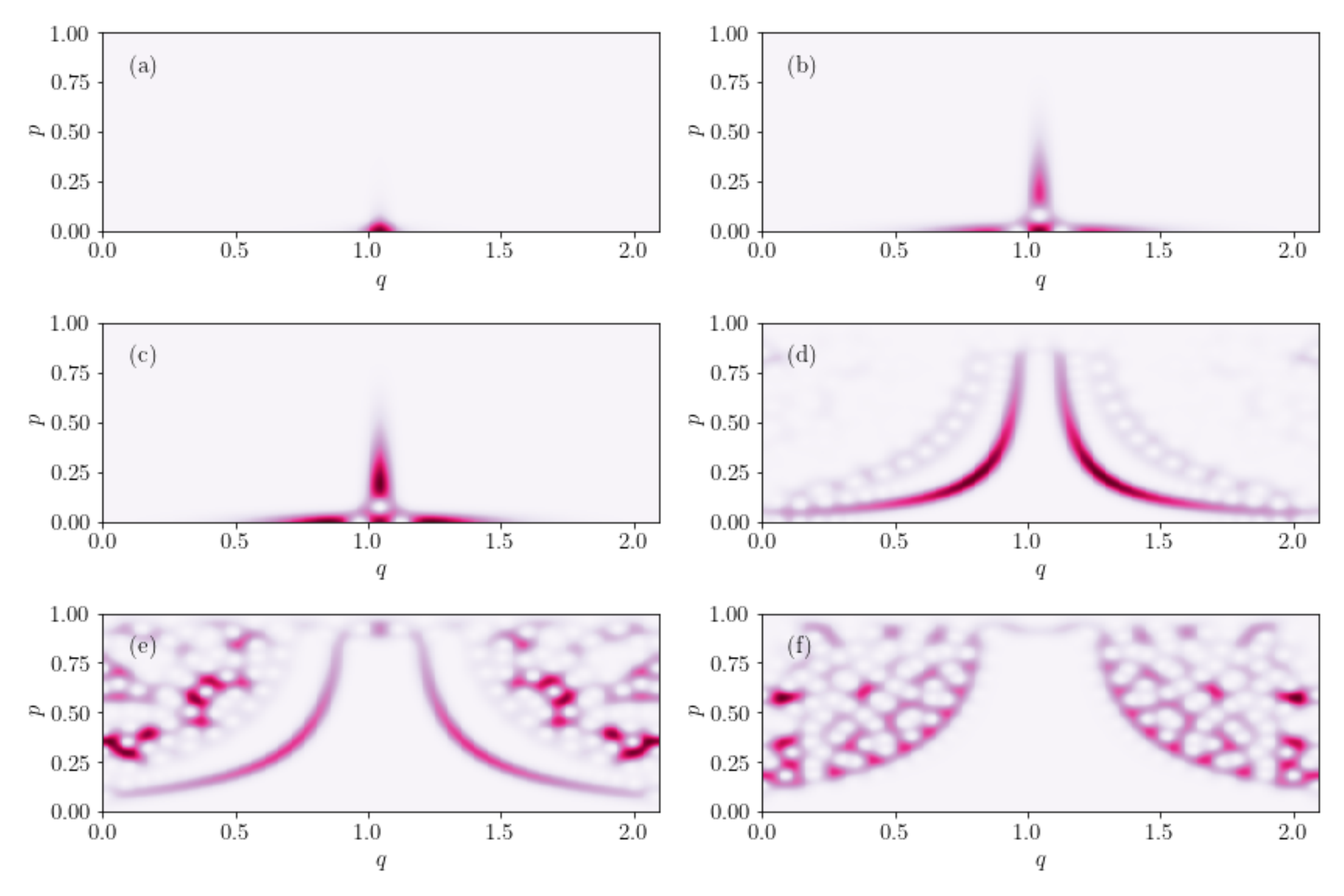}
   \par\end{centering}
 \caption{ Examples of typical Poincar\'e-Husimi functions at various
   $k$:  641.670665, 654.237553, 647.954107, 854.360748, 865.888129, and
   858.033634 in (a) to (f), respectively. Their structure is strongly
   determined by the classical stickiness structures in Figs. \ref{fig1}
   and \ref{fig2}. Due to the reflection and time reversal symmetries we plot
   only one quarter of the phase space.
   Higher color intensity corresponds to higher values of the PH function.}
  \label{fig3}
\end{figure*}

Let us define the quantum phase space. One way is to calculate the Wigner
functions \cite{Wig1932} based on $\psi_m({\bf r})$.
However, in billiards it is more natural and convenient
to calculate the Poincar\'e-Husimi (PH) functions, based
on the boundary function (\ref{BF}). The Husimi functions \cite{Hus1940} are
Gaussian smoothed Wigner functions, which makes
them positive definite. We can treat them as quasi-probability 
densities.  Following  Tualle and Voros \cite{TV1995} and B\"acker et al
\cite{Baecker2004}, we introduce \cite{BatRob2013A,BatRob2013B} 
the properly ${\cal L}$-periodized coherent states
centered at $(q,p)$, as follows

\bea \label{coherent}
c_{(q,p),k} (s) & =  & \sum_{m\in {\bf Z}} 
\exp \{ i\,k\,p\,(s-q+m\;{\cal L})\}  \times \\ \nonumber
 & \exp & \left(-\frac{k}{2}(s-q+m\;{\cal L})^2\right). 
\eea
The Poincar\'e-Husimi function is defined as the absolute square
of the projection of the boundary function $u(s)$ onto the coherent
state, namely

\be \label{Husfun}
H_m(q,p) = \left| \oint c_{(q,p),k_m} (s)\;
u_m(s)\; ds \right|^2.
\ee
All eigenstates are chaotic in the sense that the entire
classical phase space $(s,p)$ is chaotic, but not uniformly chaotic.
Namely, due to the classical stickiness regions surrounding the bouncing ball
regions (of MUPO) of Figs. \ref{fig1} and \ref{fig2}
the PH functions are localized in various
regions, as shown in Fig. \ref{fig3}: Some are strongly localized at the very
center of the bouncing ball region, some are surrounding it inside
the virtual boundary between the inner and outer part of the stickiness
region, some are localized on this boundary, some are localized
outside this boundary in a nonuniform way,
and finally some are rather uniformly spread in the outside region,
not penetrating into the inner region. All eigenstates and PH functions
have been calculated for a half billiard, for the odd parity.
In the following we perform the quantitative analysis of
the degree of localization, by calculating the  entropy
localization measure $A$.

\section{The localization measures:
  The entropy localization measure A and the normalized
  inverse participation ratio nIPR}
\label{sec4}

\subsection{The definition of localization measure}
\label{sec4.1}

The degree of localization can be quantified in at least three different
ways: entropy localization measure $A$,  correlation localization measure $C$,
and the normalized inverse participation ration  $R=nIPR$. We have shown
\cite{BatRob2013A,BatRob2013B,BLR2020}
that they are linearly related and thus equivalent.

The {\em entropy localization measure} of a {\em single
eigenstate}  $H_m(q,p)$, denoted by $A_m$ is defined as

\be \label{locA}
A_m = \frac{\exp I_m}{N_c},
\ee
where

\be  \label{entropy}
I_m = - \int dq\, dp \,H_m(q,p) \ln \left((2\pi\hbar)^f H_m(q,p)\right)
\ee
is the information entropy.  Here $f$ is the number of degrees
of freedom (for 2D billiards $f=2$, and for surface of section it is
$f=1$) and $N_c$ is a number of cells on the 
classical chaotic domain, $N_c=\Omega_c/(2\pi\hbar)^f$, where
$\Omega_c$ is the classical phase space volume of the classical chaotic component.
In the case of the
uniform distribution (extended eigenstates) $H=1/\Omega_C={\rm const.}$
the localization measure is $A=1$, while in the case of the strongest localization
$I=0$, and $A=1/N_C \approx 0$.
The Poincar\'e-Husimi function $H(q,p)$
(\ref{Husfun}) (normalized) was calculated on the grid points $(i,j)$
in the phase space $(s,p)$,  and
we express the localization measure in terms of the discretized function.
In our numerical calculations we have put $2\pi\hbar=1$, and
thus we have $H_{ij}=1/N$, where $N$ is the number of grid points,
in case of complete extendedness, while for maximal localization
we have $H_{ij}=1$ at just one point, and zero elsewhere.
In all calculations we
have used the grid of $400\times 400$ points, thus $N = 160000$.

\begin{figure}[H]
 \begin{centering}
    \includegraphics[width=9cm]{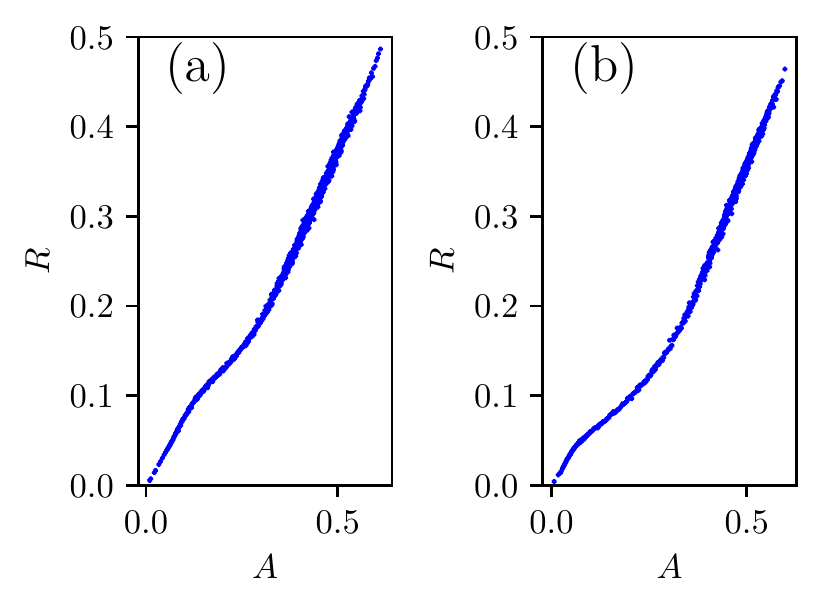}
   \par\end{centering}
 \caption{The relationship between the normalized inverse
   participation ratio $R=nIPR$ and the entropy localization
   measure $A$ averaged over the $20$ consecutive Poincar\'e-Husimi
   functions, for the $20000$ consecutive  eigenstates with $k$ above
   $k_0=640$ (a) and $k_0=2880$ (b).}
  \label{fig4}
\end{figure}

As mentioned in the introduction, the definitions of localization measures
can be diverse, and the question arises to what extent are the results
objective and possibly independent of the definition. Indeed, in reference
\cite{BatRob2013A}, it has been shown that $A$ and $C$ (based on the correlations)
are linearly related and thus equivalent. Moreover, we have introduced
\cite{BLR2019B,BLR2020} also the normalized inverse participation
ratio $R=nIPR$, defined as follows

\be \label{nIPR}
R= \frac{1}{N} \frac{1}{\sum_{i,j} H_{ij}^2},
\ee
for each individual eigenstate $m$. Here the normalization
$\sum_{ij}H_{ij}=1$ has been done.
However, because we expect fluctuations
of the localization measures even in the quantum ergodic regime
(due to the scars etc), we must perform some averaging over an
ensemble of eigenstates, and for this we have chosen $20$ consecutive
eigenstates.
The linear relation of $R=nIPR$ versus $\langle A \rangle$  has been
clearly demonstrated in Refs. \cite{BLR2019B,BLR2020} for the
stadium billiards and the mixed-type billiards (Robnik billiard,
\cite{Rob1983,Rob1984}), while in the present work for the
lemon billiard we find approximate agreement with the linear
relationship, shown in Fig. \ref{fig4}.

Also, importantly, very recently we have shown that such a
linear relationship is valid in the Dicke model. Its
classical analog based on coherent states is a
Hamilton system with a smooth potential \cite{WR2020}.
A similar finding was reported in Ref. \cite{BorIzrSanZel2016}
and references therein.
Therefore we believe that such relationship is generally true,
independent of a specific model system (billiards or smooth potentials).

In the following we shall use exclusively $A$ as the measure
of localization.

\subsection{The distributions of the localization measures  A}
\label{sec4.2}

One of the main questions addressed in this paper are the
statistical properties of $A$. In our previous works it has been
shown that in the stadium (Bunimovich billiard) $A$ obeys the
beta distribution \cite{BLR2020}, while in the mixed-type
Robnik billiard the beta distribution appears at sufficiently
uniform chaoticity and sufficiently large energies \cite{BLR2019B}.
In the case of the Dicke model
it has also been found that $A$ are distributed according to the
beta distribution \cite{WR2020}. Thus, we believe that
beta distribution of $A$ is universally valid, provided that the
stickiness regions and effects do not exist, so that we have
uniform chaoticity (constant value of $S$ in the S-plots introduced
by Lozej \cite{Lozej2020}). Our aim in the present work is
to clearly demonstrate the effects of stickiness in the quantum
properties of classically chaotic Hamilton systems, in our case the
lemon billiard B=1/2.

The so-called {\em beta distribution} is

\be  \label{betadistr}
P(A) = C A^a (A_0-A)^b,
\ee
where $A_0$ is the upper limit of the interval $[0,A_0]$ on
which $P(A)$ is defined, and the two exponents $a$ and $b$
are positive  real numbers, while $C$ is the normalization constant
such that $\int_0^{A_0} P(A)\,dA = 1$, i.e.

\be \label{C}
C^{-1} = A_0^{a+b+1} B(a+1,b+1),
\ee
where $B(x,y) = \int_0^1 t^{x-1} (1-t)^{y-1} dt$ is the beta function.
Thus we have for the first moment

\be \label{mA}
\mA = A_0 \frac{a+1}{a+b+3},
\ee
and for the second moment

\be \label{2mA}
\left< A^2 \right>  = A_0^2 \frac{(a+2)(a+1)}{(a+b+4)(a+b+3)}
\ee
and therefore for the standard deviation  $\sigma=\sqrt{ \sA -\mA^2}$

\be \label{sigmaA}
\sigma^2  = 
A_0^2 \frac{(a+2)(b+2)}{(a+b+4)(a+b+3)^2},
\ee
such that asymptotically  $\sigma \approx A_0 \frac{\sqrt{b+2}}{a}$ when
$a\rightarrow \infty$. In this limit $P(A)$ becomes Dirac delta
function peaked at $A=A_0$.

\begin{figure}[H]
 \begin{centering}
    \includegraphics[width=9cm]{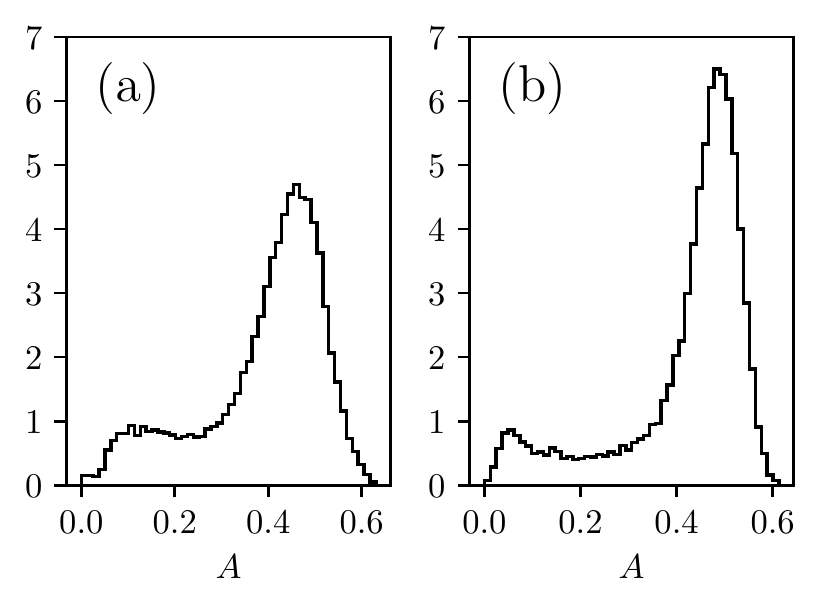}
   \par\end{centering}
 \caption{The histograms of the distribution of the entropy
   localization measure $P(A)$ for $19987$ eigenstates above
   $k_0=640$ in (a), and for $19960$ eigenstates above $k_0=2880$ in (b).
 The calculation is for the half billiard of odd parity.}
  \label{fig5}
\end{figure}  
In Fig. \ref{fig5} 
we show a selection of typical distributions $P(A)$. We clearly see
the nonuniversal bimodal distribution, deviating from the
beta distribution, and this applies to all energies considered $E_0=k_0^2$,
namely for $k_0= 640,\; 920, \; 1200, \; 1480, \; 1760, \; 2040,
\; 2320, \; 2600, \; 2880$. Therefore in  Fig. \ref{fig5}
we show only the cases
$k_0=640$ and $k_0=2880$, as in between there is no qualitative
difference. The structure is similar to the structure of $P(A)$ in
the mixed-type billiard (Robnik billiard) in the regime of
strongly nonuniform chaoticity \cite{BLR2019B}.

It should be noted that losing a few states, which
can happen, does not affect the result for $P(A)$ in any significant way.
Also, the statistical significance is very high,
which has been carefully checked by using a (factor 2) smaller
number of objects
in all histograms, as well as by changing the size of the bins.

The limiting case $a \rightarrow \infty$ in Eqs.(\ref{mA},\ref{sigmaA})
comprising the
fully extended states in the limit $\al \rightarrow\infty$ shows that the
distribution tends to the Dirac delta function peaked at $A_0$,
thus $\sigma=0$ and $P(A)=\delta(A_0-A)$, in agreement with Shnirelman's theorem
\cite{Shnirelman1974}, which is not observed in our case, but would
appear at higher energies $E=k_0^2$. In our case the
characteristic classical transport time $t_T$ cannot be uniquely defined,
as it varies widely with the location of the initial conditions
with respect to the
stickiness region. Nevertheless, we predict that for sufficiently
high energies $k_0^2$ eventually all relevant classical transport
times become sufficiently small. The Heisenberg time $t_H$
is constant, and the semiclassical parameter $\alpha$ (\ref{alpha})
for ergodic billiard is \cite{BatRob2013A,BatRob2013B}

\be \label{alphabilliard}
\alpha= \frac{t_H}{t_T} = \frac{{\cal L}\; k_0}{\pi\;N_T},
\ee
where $N_T$ is the number of collisions associated with the
transport time $t_T$.  Thus $\alpha\rightarrow \infty$ 
as $k_0\rightarrow \infty$, and we need even higher energies
to see this transition into the universal regime exhibiting
the beta distribution for $P(A)$.

\begin{figure*}
 \begin{centering}
    \includegraphics[width=1\textwidth]{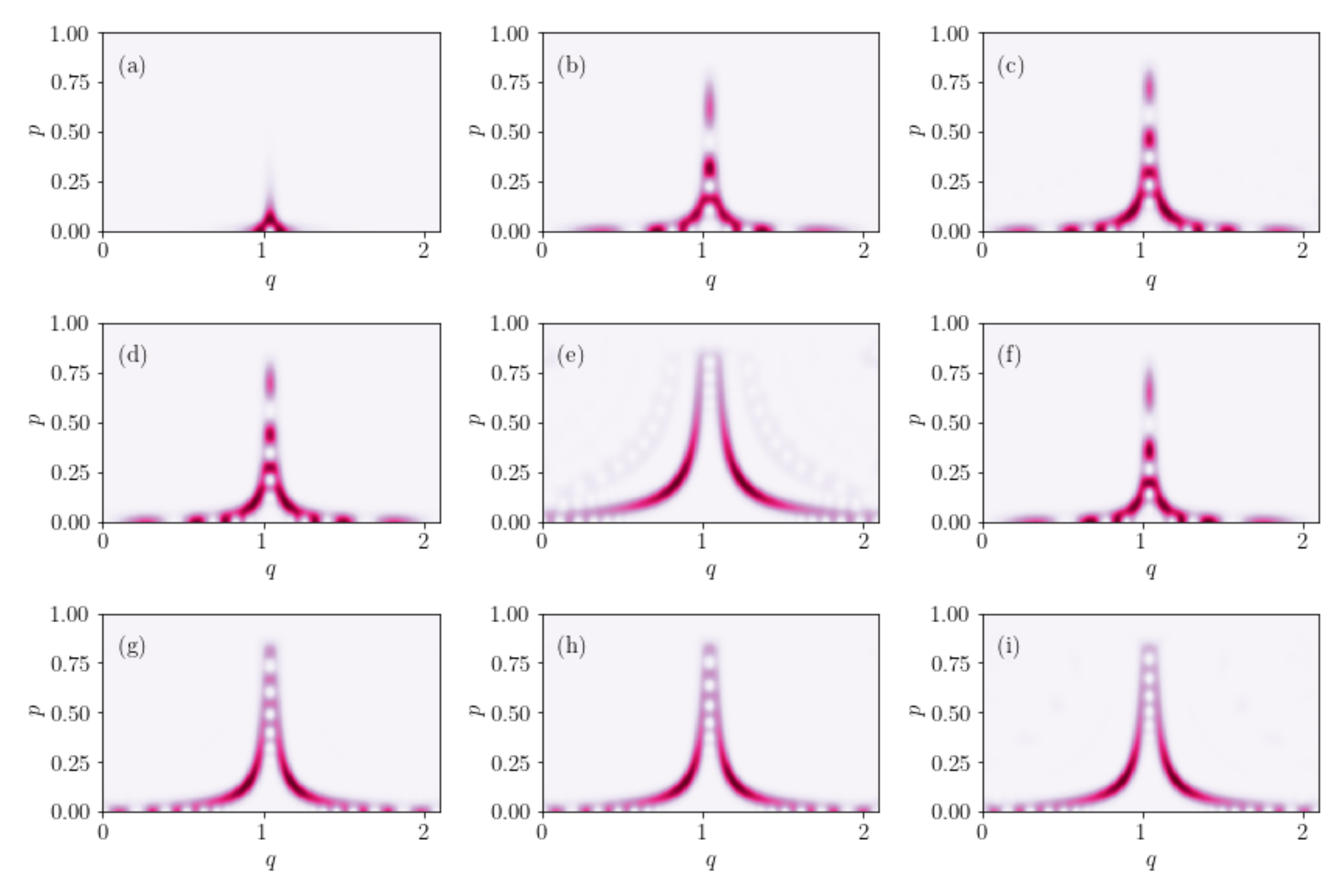}
   \par\end{centering}
 \caption{ Examples of the inner localized Poincar\'e-Husimi
   functions at various $k$: 643.241487, 646.390631, 657.395766,
   699.804421, 714.221189, 718.648216, 754.840525, 764.290106, 803.599980,
   in (a) to (f), respectively. Their structure is strongly
   determined by the classical stickiness structures in Figs. \ref{fig1}
   and \ref{fig2}. The calculation is for the half billiard of odd
 parity. Due to the reflection and time reversal symmetries we plot
 only one quarter of the phase space. Higher color intensity
 corresponds to higher values of the PH function.}
  \label{fig6}
\end{figure*}

\begin{figure*}
 \begin{centering}
    \includegraphics[width=1\textwidth]{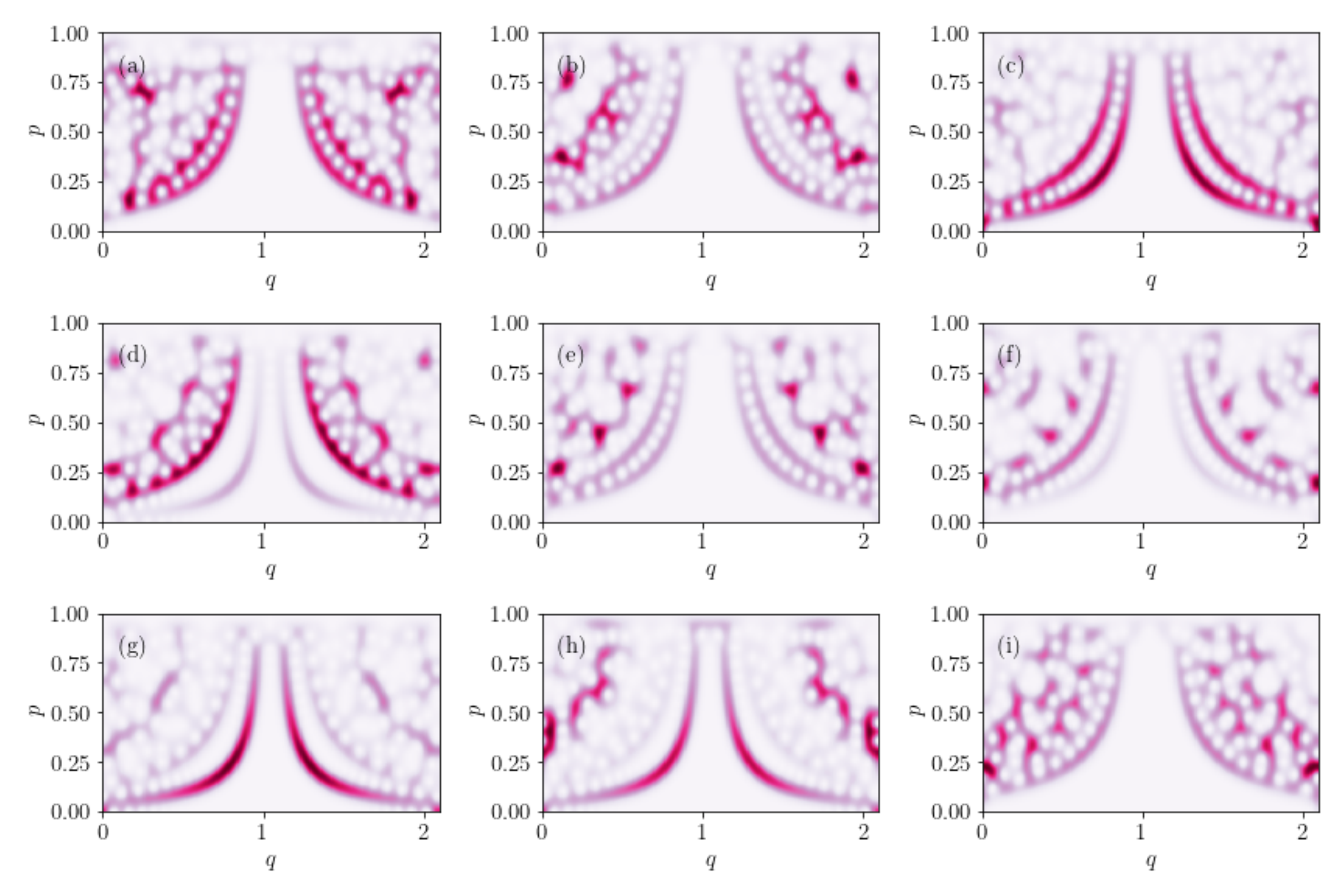}
   \par\end{centering}
 \caption{ Examples of the outer localized Poincar\'e-Husimi
   functions at various $k$: 645.621455, 645.653750, 645.989357,
   646.691491, 646.835747, 647.615914, 648.527598, 648.609436, 650.435253
   in (a) to (f), respectively. Their structure is strongly
   determined by the classical stickiness structures in Figs. \ref{fig1}
   and \ref{fig2}. The calculation is for the half billiard of even
 parity. Due to the reflection and time reversal symmetries we plot
 only one quarter of the phase space. Higher color intensity
 corresponds to higher values of the PH function.}
  \label{fig7}
\end{figure*}

We have analyzed the PH functions of the states taken from the smaller
peak around $A \approx 0.1$, and from the larger peak $A \approx 0.45$.
In the first case we see strongly localized inner states,
inside the stickiness region (Fig. \ref{fig6}), while in the second
case the states are localized outside the stickiness region,
either uniformly or nonuniformly (Fig. \ref{fig7}).

\begin{figure}[H]
 \begin{centering}
    \includegraphics[width=9cm]{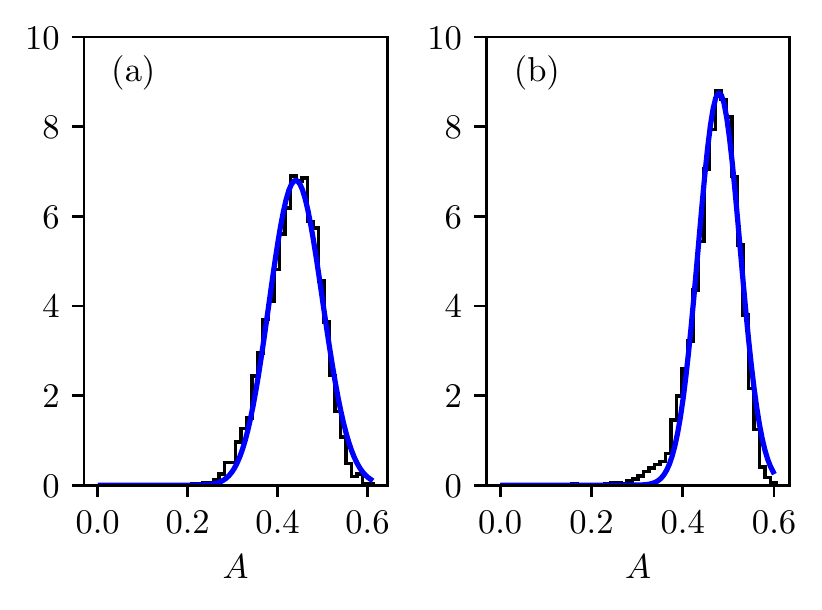}
   \par\end{centering}
 \caption{The histograms of the distribution of the entropy
   localization measure $P(A)$ for $12678$ eigenstates above
   $k_0=640$ in (a), and for $15571$ eigenstates above $k_0=2880$ in (b).
   The distributions are quite well fitted with the beta distribution
   with the parameters $(a,b)= (26.083,31.531)$ in (a) and
   $(a,b)= (47.255, 50.399)$ in (b). The states have been
   selected by the criterion of maximal overlap with the outer
 chaotic region. The calculation is for the half billiard of odd
 parity.}
  \label{fig8}
\end{figure}

If we separate the states belonging to the two peaks, using
some overlap criterion (taking only the PH functions  that
maximally overlap with the outer chaotic region),
and thus consider only the family of states
belonging to the larger peak, which "live" outside the
stickiness region, we find a unimodal distribution
which is quite well described by the beta distribution as
demonstrated in Fig. \ref{fig8}, characteristic of the
systems and regimes with no stickiness (uniform chaoticity) as
demonstrated in Refs. \cite{BLR2020,BLR2019B}.

\section{The spectral statistics:
  Berry-Robnik-Brody distribution}
\label{sec5}

Now we turn to the spectral analysis, namely the analysis of the
level spacing distribution. For an introduction see Ref. \cite{Rob2020}.
Since the billiard is ergodic,
one would expect the Brody distribution \cite{Bro1973,Bro1981},

\be \label{BrodyP}
P_B(S) = c S^{\beta} \exp \left( - d S^{\beta +1} \right), \;\;\; 
\ee
where by normalization of the total probability and the first moment we have

\be \label{Brodyab}
c = (\beta +1 ) d, \;\;\; d  = \left( \Gamma \left( \frac{\beta +2}{\beta +1}
 \right) \right)^{\beta +1}
\ee
with  $\Gamma (x)$ being the Gamma function. It interpolates the
exponential and Wigner distribution as $\beta$ goes from $0$ to $1$.
The corresponding gap probability is

\be \label{BrodyE}
{\cal E}_B(S)  =  \frac{1}{\gamma (\beta +1)  } 
  Q \left( \frac{1}{\beta +1}, \left( \gamma S \right)^{\beta +1} \right),
\ee
where $\gamma=\Gamma \left(\frac{\beta +2}{\beta +1}\right)$ 
and $Q(a, x)$ is the incomplete Gamma function

\be \label{IGamma}
Q(a, x) = \int_x^{\infty} t^{a-1} e^{-t} dt.
\ee
The degree of localization which determines $\beta$ is controlled by
the parameter $\alpha$ (\ref{alpha}). However, due to the effects of stickiness
the classical transport time cannot be defined unambigously, as it
depends strongly
on the location of the initial conditions in the phase space $(s,p)$,
and therefore so does $\alpha$ as well. Due to the strong stickiness around the
center of the bouncing ball region quantum mechanics "sees" effectively
a hole, whose size decreases with energy, and thus this hole
plays a role of a quasi-regular region (the complement of the outer
chaotic region) for most of the eigenstates,
as has been demonstrated in the PH functions.
The mechanism behind this phenomenon is the existence of a cantorus, or several
cantori, which present a border between the inner and outer region.
Namely, the holes of a cantorus are nonpermeable for the
quantum mechanics (waves) if the flux is smaller than a Planck cell,
but thus become permeable at higher energies
(or smaller Planck constant $2\pi\hbar$)
\cite{MacKayMeiss1988,CasPro1999,LozejThesis}. Therefore we must
expect that the level spacing distribution will be well described by
the Berry-Robnik-Brody (BRB) distribution, with the two parameters,
$\mu_1$ measuring the relative size of the quasi-regular region (and the
relative density of the corresponding level sequence)
and $\beta$ measuring the strength of
the localization of the chaotic part of the spectrum.

The BRB distribution is calculated as the second derivative of the
gap probabiliy  ${\cal E}$,

\be \label{BRB}
P(S)= \frac{d^2 {\cal E} }{dS^2}
\ee
where the total gap probability is the product of the regular
(Poissonian part) ${\cal E}_P=\exp (-S)$ and of the chaotic
part (\ref{BrodyE}),

\be \label{totalgap}
{\cal E} (S) = {\cal E}_P(\mu_1S)\; {\cal E}_B (\mu_2S)=
\exp(-\mu_1S)\; {\cal E}_B (\mu_2S),
\ee
where  $\mu_1+\mu_2=1$. The resulting BRB distribution captures
both effects, the quasi-divided quantum phase space, and the localization
on the outer chaotic component. 

This expectation is excellently confirmed in our numerical calculations.
It is observed that the value of $\beta$ fluctuates around the
value $0.8$, depending on the symmetry class and the energy $k_0^2$,
while $\mu_1$ decreases almost monotonically with increasing energy $k_0^2$.
At even higher energies, which we have not yet reached, $\beta$ is
expected to increase towards $1$  and $\mu_1$ to zero.
In this limit both the division of
the phase space and the localization effects disappear and we would
find just GOE level spacing distribution, well approximated by the
Wigner distribution, which is Brody distribution (\ref{BrodyP})
at $\beta=1$.

\begin{figure*}
 \begin{centering}
    \includegraphics[width=12cm]{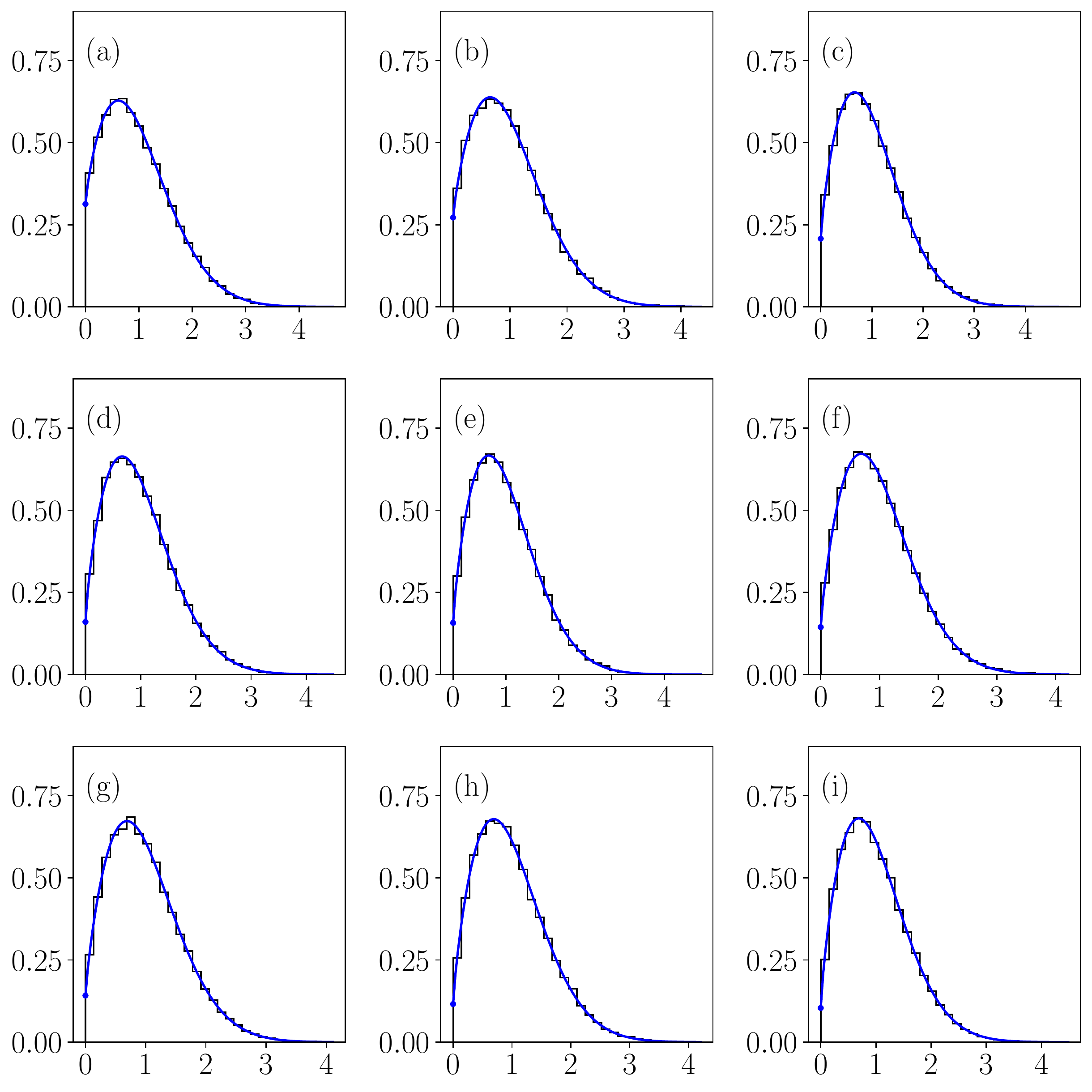}
   \par\end{centering}
 \caption{The histograms of the level spacing distribution $P(S)$ for
 nine energy intervals each starting at 
$k_0= 640,\; 920, \; 1200, \; 1480, \; 1760, \; 2040,
\; 2320, \; 2600, \; 2880$ and comprising about 40000 levels
that include all four symmetry groups (about 10000 levels of
each symmetry group) for each $k_0$.
The fitting parameters $(\beta,\mu_1)$
are from (a) to (i):  (0.827, 0.171), (0.844, 0.147), (0.806, 0.110),
(0.778, 0.083), (0.799, 0.082), (0.813, 0.075), (0.816), 0.074),
(0.801, 0.060), (0.789, 0.053). By the thick dot we denote the
value of $P(S=0)$, which decreases monotonically with increasing
$k_0$.}
  \label{fig9-1}
\end{figure*}

In Fig. \ref{fig9-1} we show the level spacing distributions
for nine energy intervals each starting at 
$k_0= 640,\; 920, \; 1200, \; 1480, \; 1760, \; 2040,
\; 2320, \; 2600, \; 2880$ and comprising about 40000 levels
that include all four symmetry groups (about 10000 levels of
each symmetry group). They are all very well fitted by the
Berry-Robnik-Brody distribution  (\ref{BRB}). Note that
the value of $P(S=0)$ monotonically decreases with increasing
energy $k_0$, as predicted: At higher energies the quantum
resolution of the classical structures in the phase space
increases, therefore the eigenstates tend towards the
ergodic regime, in which the stickiness plays lesser and lesser
role ($\mu_1$ and $P(S=0)$ tend to zero).

\begin{figure*}
 \begin{centering}
    \includegraphics[width=1\textwidth]{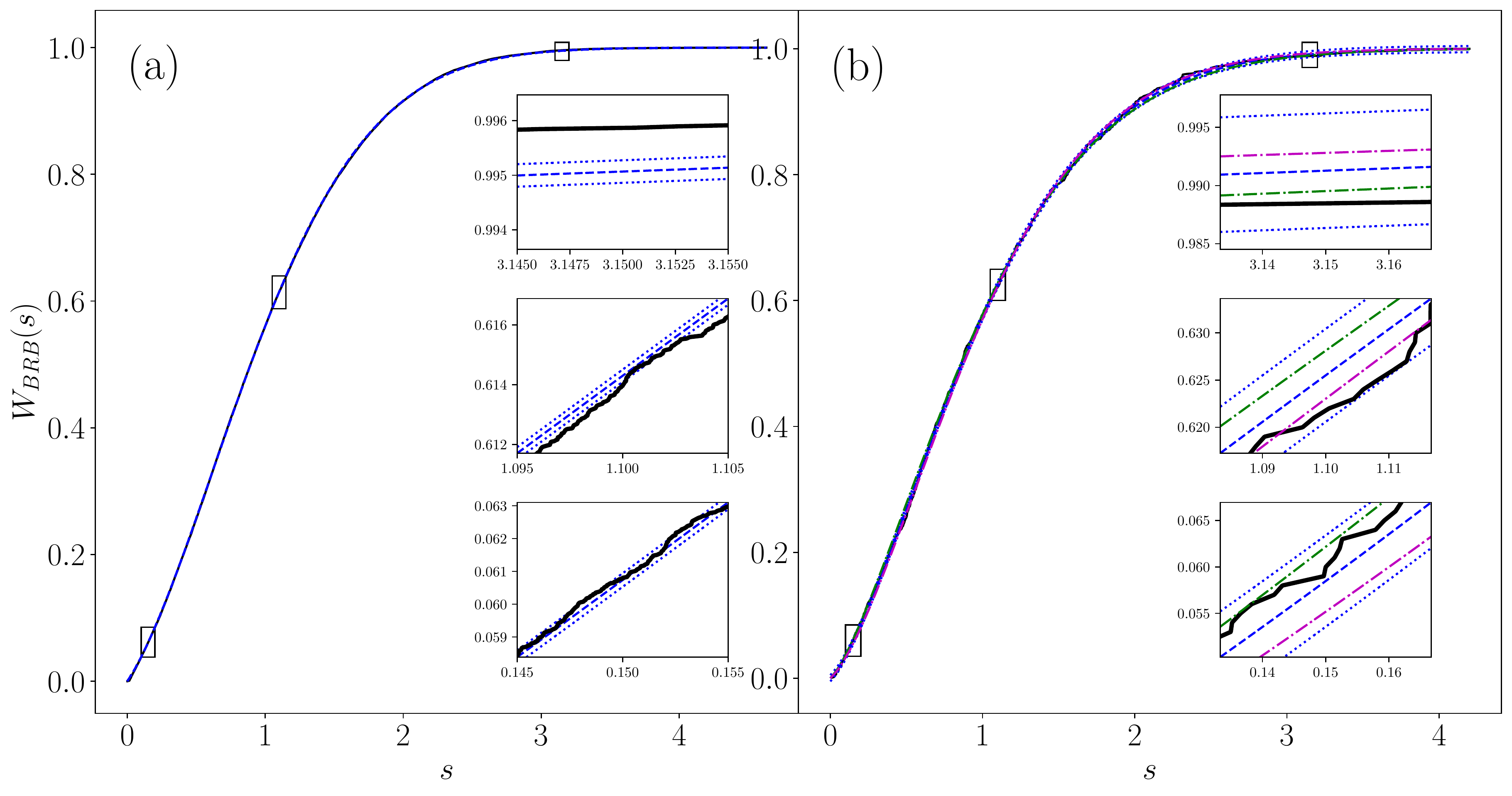}
   \par\end{centering}
 \caption{The cumulative level spacing distribution $W(S)$ for
 two energy intervals each starting at 
 $k_0= 640$: in (a) 39965 levels comprising all four parities,
 and in (b), 1000 levels of odd-odd parity. The parameters
 $(\beta,\mu_1)$ are (0.827, 0.171) in (a) and (0.569, 0.092)
 in (b). To display small
 deviation of data from the best fitting BRB distribution
 we show in the insets magnification: The thick lines (black)
 are the numerical data, the best fitting BRB curve is dashed
 (blue), the dotted (blue) lines designate the $\pm$ one standard
 deviation from the best fitting BRB curve, and the 
 dash-dotted lines (magenta and green)
 denote the BRB curves with the same $\mu_1$ but different
 $\beta$ by the amount $\pm 0.05$. One should observe 
 the significantly different values of $\beta$ and $\mu_1$
compared  between (a) and (b),
 showing that the statistics based on almost 40000 levels
(a) is better than in the case of only 1000 levels (b).}
  \label{fig9-3}
\end{figure*}

%
%
%
%
%
%
%
In order to verify the goodness of the BRB
distribution we plot in Fig. \ref{fig9-3} also the cumulative
level spacing distribution for the case $k_0=640$, twofold,
for the 1000 consecutive eigenstates of the odd-odd parity in (b),
and for the 39965 eigenstates of all four parities in (a).
We see very good agreement. It is
seen that increasing the energy range and the number of
levels  significantly changes the values of $\beta$ and $\mu_1$
and the quality of the  theoretical fitting BRB, although
two effects work against each other:
Increasing the energy range makes $\beta$ less sharply defined
while increasing the number of objects decreases the
statistical error. We may conclude that the agreement is
excellent.

\begin{figure}[H]
 \begin{centering}
    \includegraphics[width=9cm]{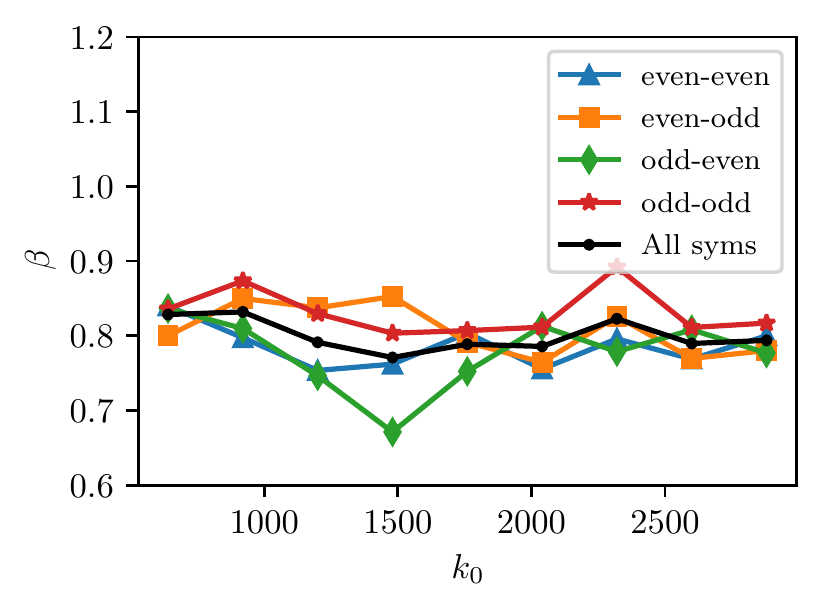}
   \par\end{centering}
 \caption{The dependence of the $\beta$ parameter on the
   energy $k_0$.  For each $k_0$ we have taken about 10000
   states of given parity above $k_0$, and also show the data
   for the ensemble of all four parities. The value of
 $\beta$ fluctuates around $\beta\approx 0.8$.}
  \label{fig10}
\end{figure}

In Fig. \ref{fig10} we show the dependence of $\beta$
on the energy $k_0$ for about 10000 levels of each
parity, and the collection of all four parities. It is
seen that $\beta$ fluctuates around $0.8$. At still higher energies
it is predicted to increase towards the value $\beta=1$
(Wigner distribution, which is 2-dim GOE),
in the deep semiclassical limit.

\begin{figure}[H]
 \begin{centering}
    \includegraphics[width=9cm]{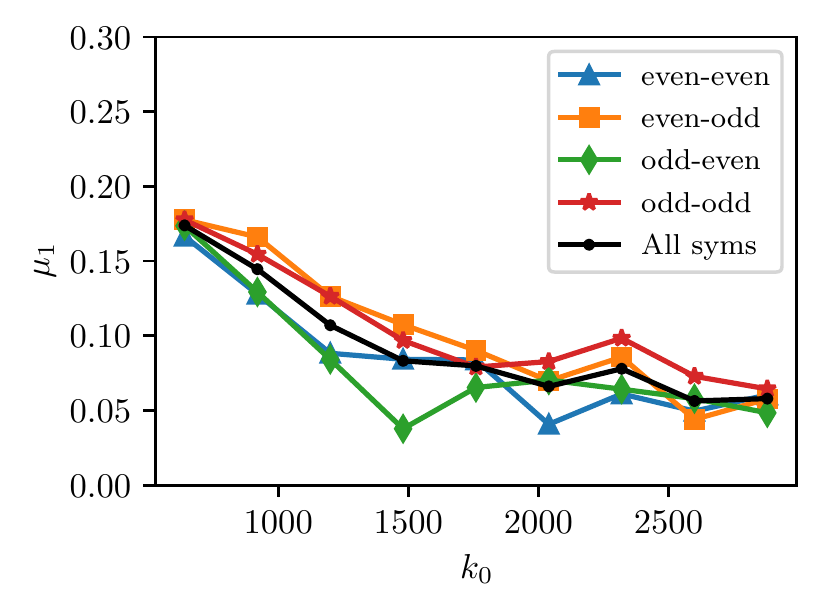}
   \par\end{centering}
 \caption{The dependence of the $\mu_1$  parameter on the
 energy $k_0$. For each $k_0$ we have taken about 10000
   states of given parity above $k_0$, and also show the data
   for the ensemble of all four parities. The value
 of $\mu_1$ decreases almost monotonically with $k_0$.}
  \label{fig11}
\end{figure}

In Fig. \ref{fig11} we show the dependence of the
parameter $\mu_1$ on the energy $k_0$, for about 10000 levels of each
parity, and the collection of all four parities, clearly showing
that it almost monotonically decreases with $k_0$. Asymptotically
it must tend to zero, as the system is (practically) ergodic.

\section{Additional comments}
\label{sec6}

In the course of our present work we have widely explored the
PH functions and the level spacing statistics by varying
all possible parameters, like $k_0$, the number of levels $n$
above $k_0$ from 1000 to 10000 taken in histograms and
cumulative level spacing ditribution, the four parities,
the size of the bins in histograms, etc. Hundreds of PH
functions have been produced and analyzed, as well as hundreds
of level spacing statistics, from plentiful different
points of view.

The general conclusion
is that the determination of the Berry-Robnik-Brody distribution
is far from trivial and the values of the parameters
$\beta$ and $\mu_1$ depend quite sensitively on the
above mentioned parameters. Of course, the most reliable data
are the largest ones, comprising typically 10000 levels
per $k_0$ and parity, on which our conclusions are based.

Another remark concerns the classical transport time $t_T$
or $N_T$ (the number of collisions associated with $t_T$),
which enters in the general expression for $\alpha$ in (\ref{alpha}),
and for a general ergodic billiard in (\ref{alphabilliard}).
This time scale cannot be uniquely defined, as its value
depends strongly on the initial conditions. Nevertheless,
a rough estimate has been done for initial conditions close
to $p=0$ (remember, the line $(s,p=0)$ is invariant) with the
result $N_T \approx 1000$.  In our case ${\cal L}= 4\pi/3=4.188790$,
therefore $\alpha \approx 4 k_0/3000$, and in the range
$k_0 \in [640,2880]$ we have  $\alpha\in [0.853, 3.84]$.
This means that we are just in the middle of the
localization transition region from $\alpha \ll 1$ to
$\alpha \gg 1$, indicating that we should see quite
strongly expressed localization of PH functions, which is
indeed the case. However, our energy interval $E = k_0^2$  with
$k_0\in[640,2880]$ is too narrow to observe the variation
of $\beta$ with $k_0$,
that is why $\beta$ fluctuates around $0.8$.

More precise estimates must be done by analyzing in
detail the structure of the strong stickiness region,
which implies at least two quite different time scales:
One inside the sticky region, and the other outside.
Quantum mechanically the boundary between them depends
on the energy $k_0^2$.
Such a more detailed analysis is left for the future.

\section{Discussion and conclusions}
\label{sec7}

We have presented the semiempirical analysis of the
chaotic ergodic lemon billiard (B=1/2), classically and
quantally. The existence of strong stickiness regions around the
invariant zero-measure bouncing ball lines as quantified
by the phase portraits (density plots) and Lozej's S-plots
has important consequences for the quantum mechanics
of the same billiard. The Poincar\'e-Husimi (PH) functions
are strongly localized and their entropy localization
measure $A$ has a bimodal distribution, qualitatively due
to the existence of basically two PH functions populations,
namely the inside ones and the outside ones. If we eliminate
the inner eigenstates, we find that $A$ obeys quite well
the beta distribution characteristic for the uniform
chaoticity (no stickiness in chaotic region and $S=1$).
The existence of such a strong stickiness region is manifested
also in the energy spectral statistics. As the quantum
mechanics "sees" the inner region, at given energy $k_0^2$,
effectively as a separate regular region as the complement
of the outer chaotic region, the level spacing distribution
is Berry-Robnik-Brody (BRB) with two parameters: $\beta$
measures the degree of localization and the level repulsion
effect, and $\mu_1$ measures effectively the size of the inner
sticky region. The agreement of data with BRB is excellent.
As $\alpha$ is roughly within the interval $\alpha\in [0.853, 3.84]$,
we see that $\beta$ is hardly changing with the energy $k_0^2$,
and fluctuates around $0.8$, while the parameter $\mu_1$
decreases almost monotonically with $k_0$, as predicted:
The quantum resolution of the classical phase space structures
increases with increasing energy. Asymptotically, when
$k_0\rightarrow \infty$
we predict $\beta \rightarrow 1$ and $\mu_1\rightarrow 0$.
However to reach these higher energies, a major
computational effort is necessary.  A more detailed
analysis of the structure of the stickiness region, the
associated transport time scales and their quantum
implications are left for the future, which requires
calculation of eigenstates and PH functions at much higher energies.

Another still open problem is the theoretical explanation of
the Brody level spacing distribution even in the case of
uniformly chaotic (no stickiness) regime with localized
PH functions. This includes the distribution of the
entropy localization measure $A$ as beta distribution.

These aspects have been explored and demonstrated
in Refs.
\cite{BatRob2013A,BatRob2013B,BLR2018,BLR2019B,BLR2020} in other billiards,
and very recently by Wang and Robnik \cite{WR2020} also
for the Dicke model, whose classical counterpart based
on the coherent states is a Hamilton system with a smooth
potential, which corroborates our findings. It seems that 
a semiclassical method based on Gutzwiller's periodic orbit theory
\cite{Gutzwiller1967,Gutzwiller1969,Gutzwiller1970,Gutzwiller1971,
Gutzwiller1980}
might be an appropriate approach \cite{Stoe,Haake}
to solve this problem.

\section{Acknowledgement}

This work was supported by the Slovenian Research Agency (ARRS) under
the grant J1-9112.

\providecommand{\noopsort}[1]{}\providecommand{\singleletter}[1]{#1}%

\end{document}